\documentclass[12pt]{article}

\usepackage{amsmath}
\usepackage{amssymb}
\usepackage{amsfonts}
\usepackage{graphicx}
\usepackage{latexsym}

\begin{document}

\title{Progress in the mathematical theory of quantum disordered systems}
\author{Walter F. Wreszinski\\
Instituto de F\'{i}sica\\
Universidade de S\~{a}o Paulo\\
Caixa Postal 66318\\
05314-970 S\~{a}o Paulo, SP , Brazil}

\maketitle

\begin{abstract}
We review recent progress in the mathematical theory of quantum
disordered systems: the Anderson transition , including some joint work with Domingos
Marchetti, the (quantum and classical) Edwards-Anderson (EA) spin-glass
model and return to equilibrium for a class of spin-glass models,
which includes the EA model initially in a very large transverse
magnetic field. 
\end{abstract}

\textbf{1 - Introduction}

In recent years there has been a significant progress in the
mathematical theory of (quantum) disordered systems. Our purpose here
is to present the main ideas (without proofs), with a clear discussion of
the ideas, concepts and methods. Some new remarks and results pertaining to sections
3 and 4 are also included. 

We shall be interested in properties of disordered systems at low
temperatures, i.e., near the ground state (absolute temperature $T =
0$); in the critical region of spin glasses, there exist the
spectacular recent rigorous results on the mean field theory (see the
review by F. Guerra \cite{Gue}), and various results for other
disordered systems, including the well understood high temperature
phase in spin glasses, are discussed in the comprehensive book by
A. Bovier \cite{Bo}. 

Our restriction to very low temperatures implies, of course, that we
shall be dealing exclusively with quantum systems. In particular, the
Ising model, when it appears, must be regarded as the anisotropic
limit of quantum (spin) systems: the fact that this is not only
mathematically so is emphatically demonstrated by the well-known fact
that the critical exponents of the Ising model in three dimensions
(see, e.g., \cite{Zinn}) are surprisingly close to those measured in
real magnetic systems, precisely because most of the latter are highly
anisotropic.  

Three important issues appearing in the above-mentioned context are
\textbf{localization}, arising in connection with the Anderson
transition, which we discuss in section 2, reporting on several works, as well as
on some joint work with Domingos Marchetti (\cite{MW},\cite{MWB}), \textbf{frustration},
which appears in short-range spin glasses such as the Edwards-Anderson
(EA) model, discussed in section 3, and the (thermodynamic version of)
\textbf{instability of the first and second kind}, which relates to
the return to equilibrium for special initial states and probability
distributions in a class of models (including the EA model) in section
4.

\textbf{2a - The Anderson transition - Generalities and review of some results}

The breakdown of translation invariance (or, more generally, Galilean
invariance in many-body systems with short-range forces) leads to the
existence of \textbf{crystals}, and the corresponding Goldstone
excitations are phonons \cite{Swi}. On the other hand, the discrete
translation invariance subgroup of the crystal is also frequently
broken by e.g. impurities. This fact brings about a number of
important new conceptual issues, and is usually modelled by the
introduction of a random local potential in a tight-binding model (for
the latter see \cite{FeIII}). The associated physical picture consists
of lightly or heavily doped semiconductors (e.g. Si doped with a
neighboring element which contributes excess electrons, for instance
P). This is the \textbf{Anderson model} (\cite{An1}, \cite{An2}),
described by the Hamiltonian 

\begin{eqnarray*}
H^{\omega }=\Delta + V^{\omega }  
\end{eqnarray*}
$$\eqno{(1)}$$
on 
$l^{2}(\mathbf{Z}^{d})$
where $\Delta$ is the (centered discrete Laplacian)
\begin{eqnarray*}
\left( \Delta u\right) _{n}=\sum_{n^{\prime }:\left\vert n-n^{\prime
}\right\vert =1}u_{n^{\prime }} 
\end{eqnarray*}
$$\eqno{(2)}$$
plus a perturbation by a random potential 
\begin{eqnarray*}
\left( V^{\omega }u\right) _{n}=V_{n}^{\omega }u_{n}
\end{eqnarray*}
$$\eqno{(3)}$$
where $\left\{ V_{n}^{\omega }\right\} _{n\in \mathbf{Z}^{d}}$ is a family
of independent, identically distributed random variables (i.i.d.r.v.) on a
probability space $\left( \Omega ,\mathcal{B},\mu \right) $, with a common
distribution $F(x)=\mu \left( \left\{ \omega :V_{n}^{\omega }\leq x\right\}
\right) $; $V_{n}^{\omega}$ is assumed to depend linearly on a quantity $v > 0$,
which is the disorder parameter, also
called coupling 
constant. The spectrum of $H^{\omega }$ is, by the ergodic theorem \cite{KuS}, almost
surely a nonrandom set $\sigma(H^{\omega })=[-2d,2d]+ \mbox{ supp } dF$
. Anderson conjectured that there exists a critical coupling
constant $0< v _{c}<\infty $ such that for $v \geq v _{c}$
the spectral measure of $H$ is pure point (p.p) for $\mu $--almost
every $\omega $, while, for $ v < v _{c}$ the spectral measure of $%
H^{\omega }$ contains two components, separated by so called
\textbf{mobility edge} $E^{\pm }$: if $E\in \lbrack E^{-},E^{+}]$
the spectrum of $H^{\omega }$ is pure absolutely continuous (a.c); in the
complementary set $\sigma (H^{\omega })\backslash \lbrack E^{-},E^{+}]$, $%
H^{\omega }$ has pure point spectrum - leading to the new important phenomenon of
\textbf{localization}, first proved for $d=1$ in \cite{GMP}, and for
$d \ge 3$ for the first time in by \cite{FMSS}, based on previous
seminal work by \cite{FS}, for fixed energy and large disorder, or for
fixed disorder and large energy (the latter meaning near the edges of
the band, i.e, the boundary of $\sigma(H^{\omega})$); see also
\cite{AM} for a much simpler proof by an entirely different method,
and references given there. 

The references given above are, of course, only symbolic for the extensive literature
on the Anderson model, but, fortunately, an excellent review on the subject, 
up to about 2007, is available \cite{J}, to which we refer for a readable and enlightening
discussion, as well as references. Our review on the Anderson model will be mostly restricted
to later developments, related, in particular, to the elusive existence of absolutely 
continuous (a.c.) spectrum and the Anderson transition. Even with this restriction, our exposition
will not be exhaustive, but we shall try to explain some of the key issues, with the aim of 
stimulating further research in this exciting area.

The first model for which a sharp metal-insulator transition was found is the one described
by the almost Mathieu operator $H_{\omega,\lambda,\theta}$ acting on $l^{2}(\mathbf{Z})$ and
given by
$$
(H_{\omega,\lambda,\theta} \Psi)(n) = \Psi(n+1) + \Psi(n-1) + f(\omega n+\theta) \Psi(n)
\eqno{(0)}
$$
with $f(\theta) = \lambda \cos(2\pi\theta)$. This model was extensively studied in the physics
and mathematics literature since the seventies, and displays, for almost every $\omega \in \mathbf{R}$,
$\theta \in \mathbf{R}$, a transition from a set of delocalized states (''metal''), characterized
by a purely a.c. spectrum for $\lambda < 2$, to a set of localized states (''insulator''), characterized
by a purely pure point (p.p.) spectrum with exponentially decreasing eigenfunctions (see \cite{J},3.4)
for a rather complete review and references). The one-dimensional nature of the model, as well as its
physical range of applications (two-dimensional electron subject to a perpendicular magnetic field,
integer quantum Hall effect) distinguish it from the Anderson model (see, again, \cite{J}, pg. 632 and
references).

A first direction in the search of a.c. spectrum led to the study of models with randomly decaying 
potentials, where $V^{\omega}_{n}$ in (3) is replaced by $a_{n}V^{\omega}_{n}$, with $a_{n} \to 0$ as
$|n| \to \infty$. For dimension $d = 1$, many sharp results were obtained, for $d \ge 2$, see \cite{Kri1},
\cite{BoSa}, and the lecture note \cite{Bou}. For the special ''random surface case'', where $V^{\omega}_{n} = 0$
outside a ''hypersurface'' (of some thickness) in $\mathbf{Z}^{d}$, see \cite{JaLa1}. In a different, but related,
approach \cite{KiSt} studied the eigenvalue statistics of a class of matrices (CMV matrices, see \cite{Da}) with
random decaying coefficients, and \cite{Sche}, random band matrices, with entries vanishing outside a band of
finite width around the diagonal, with a view to determining the crossover between an ''insulating'' regime
with localized eigenfunctions and weak eigenvalue correlations and a ''metallic'' regime with extended 
eigenfunctions and strong eigenvalue repulsion.

Much more recently, Krishna \cite{Kri2} presented a class of Anderson type operators with independent, non-stationary
(non-decaying) random potentials, with pure a.c. spectrum in the middle of the band for small disorder, and
\cite{Kri3} a class of continuous, non-decaying and non-sparse potentials exhibiting a spectral transition. These
papers also open interesting possibilities of applications of well-known methods, powerful in other contexts
(positive commutators, virial theorem, Mourre theory) in the context of the Anderson transition.

In the nest section 2b, we confine our attention to two special models of the Anderson transition. One is the 
Anderson model on the Bethe lattice, first sudied by A. Klein in a seminal paper \cite{Kl} (see also \cite{J} for
a discussion and further references). This work was further considerably extended by M. Aizenman and S. Warzel
(\cite{AiWa1}, \cite{AiWa2}, \cite{AiWa3}, \cite{AiWa4}, \cite{AiWa5}). The other is a sparse model in $d \ge 2$ 
dimensions, studied by D. Marchetti and the present author \cite{MW}, \cite{MWB}. Our special emphasis on these two
models is justified by the fact that they are amenable to a quite detailed analysis ; it does not, of course, imply 
any comparative judgment og value regarding the other approaches mentioned in section 2a. Although, as we shall see, 
both models have serious shortcomings when compared with the original proposal (1)-(3), both yield a pallette of 
stimulating insights. We treat the sparse model first, because it leads to the introduction of some notions which
are used elsewhere in the sequel.

\textbf{2b1-Sparse models}

Another collection of models introduced to study the Anderson  transition have $V^{\omega}_{n}$ in (3) of the form
\begin{eqnarray*}
V_{n}^{\omega }=\sum_{i}\varphi _{i}^{\omega }(n-a_{i})~,  
\end{eqnarray*}
with elementary potential ('' bump") $\varphi ^{\omega }:
\mathbb{Z}^{d}\longrightarrow \mathbb{R}$ satisfying a uniform integrability
condition
\begin{eqnarray*}
\left\vert \varphi ^{\omega }(z)\right\vert \leq \frac{C_{0}}{1+\left\vert
z\right\vert ^{d+\varepsilon }}
\end{eqnarray*}
for some $\varepsilon >0$ and $0<C_{0}<\infty $ and
\begin{eqnarray*}
\lim_{R\rightarrow \infty }\frac{\#\left\{ i:\left\vert a_{i}\right\vert
\leq R\right\} }{R^{d}}=0\ .  
\end{eqnarray*}
$$\eqno{(4)}$$
Due to condition of zero concentration, potentials such as $V$ 
are called sparse and have been intensively studied in recent years since
the seminal work by Pearson \cite{Pe1} in dimension $d=1$ by
Krishna \cite{Kri4}, Krutikov \cite{Kru} and Molchanov \cite{Mo}, \cite{Mo1} in the 
multidimensional case.  For $d\geq2$ the interaction between bumps is weak 
while for $d=1$ the phase
of the wave after propagation between distant bumps becomes 
''stochastic". We shall report on a version of these models recently studied
by D. Marchetti and the present author \cite{MW}, \cite{MWB}, for which a rather
detailed information is available. We consider (infinite) Jacobi matrices 
\begin{eqnarray*}
(J_{0}u)_{n}=u_{n+1}+ u_{n-1}\;,
\end{eqnarray*}
with the perturbation potential
$$(V^{\omega}u)_{n} = v^{\omega}_{n} u_{n} $$
with $u=(u_{n})_{n\geq 0}\in l_{2}(\mathbb{Z}_{+})$ ,and
\begin{eqnarray*}
v_{n}^{\omega}=\left\{
\begin{array}{lll}
v & \mathrm{ if } & n=a_{j}^{\omega }\in \mathcal{A}\,, \\ 
0 & \mathrm{ if } & \mbox{ otherwise }\,,
\end{array}
\right.  
\end{eqnarray*}
for $v\in (0,1)$. 
$\mathcal{A}=\{a_{j}^{\omega }\}_{j\geq 1}$ denotes a random set of natural numbers
$a_{j}^{\omega }=a_{j}+\omega _{j}$ ,where $a_{j}$ satisfy the 
''sparseness condition'':
\begin{eqnarray*}
a_{j}-a_{j-1}=\beta ^{j}\;,\qquad \qquad j=2,3,\ldots 
\end{eqnarray*}
with $a_{1}+1=\beta \geq 2$ and $\omega _{j}$, $j\geq 1$,
are independent r.v. on a probability space
 $(\Omega ,\mathcal{B},\mu )$ uniformly distributed on a set
 $\Lambda_{j}=\{-j,\ldots ,j\}$. These variables introduce an
 uncertainty in the positions of 
those points for which $v_{n} \ne 0$: such models are called
\textbf{Poisson models} (see \cite{J} 
and references given there). Note that the support of the $\omega
_{j}$ only grows linearly with the suffix $j$. 
We write $J^{\omega} = J_{0} + V^{\omega}$, and denote by
 $J_{\phi }^{\omega }$ the operator associated to $J^{\omega }$ 
on the Hilbert space
 $\mathcal{H}$ of the square integrable sequences $u=\left(
u_{n}\right) _{n\geq -1}$ which satisfy a $\phi $-b.c. at $-1$:

\begin{eqnarray*}
u_{-1}\cos \phi -u_{0}\sin \phi =0  
\end{eqnarray*}
 
The essential spectrum of $J_{\phi}^{\omega}$ equals $[-2,2]$: it will
be represented as  
$\lambda = 2\cos \alpha$ with $\alpha \in [0,\pi)$. Zlatos \cite{Zl}
proved that this model exhibits a sharp 
transition from s.c. to p.p. spectrum. This was shown independently in
(\cite{CMW1},\cite{MWB}):  

\textbf{Theorem 1} Let $J_{\phi }^{\omega}$ be as above. Let
\begin{eqnarray*}
I\equiv \left\{ \lambda \in \lbrack -2,2]\backslash 2\cos \pi \mathbb{Q}:
\frac{1}{v^{2}}(\beta -1)(4-\lambda ^{2})>1\right\}
\end{eqnarray*}
with $v\in (0,1)$ and $\beta \in \mathbb{N}$, $\beta \geq 2$ . Then,
for almost all 
$\omega $ with respect to the uniform product measure on 
$\Lambda =\times_{j=1}^{\infty }\left\{ -j,\ldots ,j\right\} $,

(a) there exists a set $A_{1}$ of Lebesgue measure zero such that the
spectrum restricted to the set $I\backslash A_{1}$ is purely singular
continuous,

(b) the spectrum of $J_{P,\phi }$ is dense pure point when restricted
to  \hfill \break $I^{c}=\left( [-2,2]\backslash 2\cos \pi \mathbb{Q}\right)
\backslash I$ for almost every $\phi \in (0,\pi )$, where $\phi $
characterizes the boundary condition. Thus it is \textbf{purely}
p.p. in this interval.

\begin{figure}[h]
\centering
\includegraphics{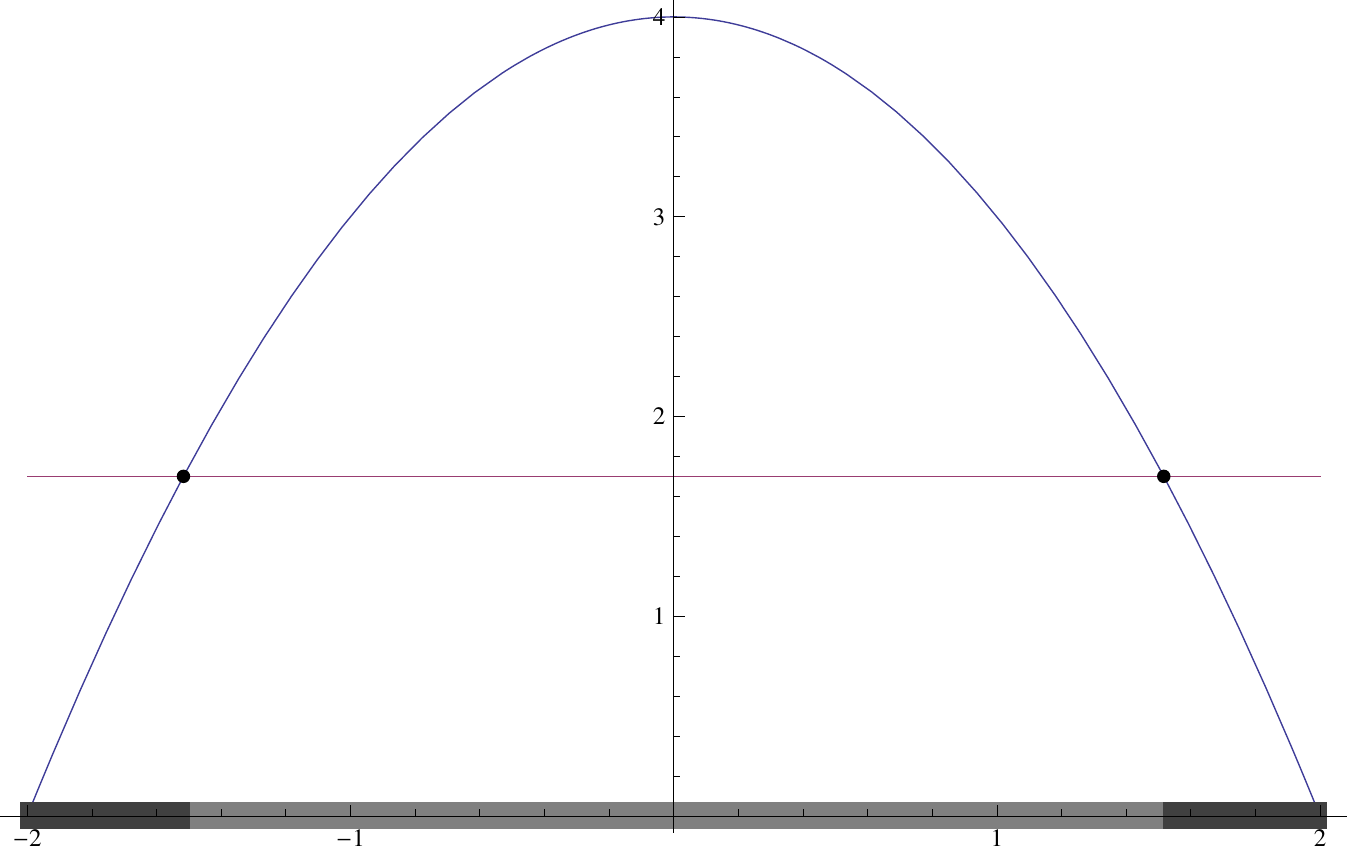}
\caption{Figure 1: In the above figure, we see the s.c. (light gray) and p.p. (dark grey)
spectra separated by the ''mobility edges''  
$\lambda^{\pm} = \pm 2 \sqrt(1-\frac{v^{2}}{v_{c}})$, with
$\frac{v}{v_{c}}=1,3038 \cdots$.} 
\end{figure}

\textbf{Remark 1} 

The basic emphasis on one-dimensional (sparse) models has an important
technical reason: the profound approach of Gilbert and Pearson
connecting the space asymptotics of eigenfunctions of the restrictions
of a large class of Sturm-Liouville operators to finite intervals to
the spectral theory of these same operators in infinite space through
the concept of \textbf{subordinacy} \cite{GP} is only available in one
dimension. Using the important transfer matrix version of this theory
due to Last and Simon \cite{LaSi}, the surprise is that, even in a
regime of strong sparsity, a spectral transition from s.c. to
p.p. spectrum (first shown by Zlatos \cite{Zl}) may be proved: this is
theorem 1. Our proof (theorem 4.7 of \cite{MWB}) differs from that in
\cite{Zl} by the use of the (optimal) metrical version of Weyl's
theorem on uniform distribution \cite{KN} (in this case, of the
so-called Pr\"{u}fer angles, see \cite{KLS}) due to Davenport,
Erd\"{o}s and LeVeque \cite{DEL}. The latter permits an
\textbf{explicit} characterization of the exceptional set in which the
spectrum may not be of p.p. type (this exceptional set arises in
connection with the concept of essential support of a measure
\cite{GP}), which reveals it to be also of p.p. type. As a result, the
spectrum at the edges of the band is proved to be \textbf{purely}
p.p.. 

The robustness of the transition depicted in theorem 1 follows from
\cite{dR}, because the Hausdorff dimension of the s.c. spectrum may be
seen to be nonzero (\cite{Zl}, \cite{CMW3}).  

Finally, the difference Laplacean (2) leads to tunneling and bands
\cite{FeIII}, with a pure a.c. spectrum. Starting from the difference
Laplacean, the figure shows that any strength of the disorder $v$,
however slight, yields p.p. spectrum in a nonempty range of energies.
In addition, this occurs in a regime of strong sparsity. This is also true
for the forthcoming models in $d \ge 2$ dimensions, and is thus a concrete
manifestation of the \textbf{tunneling instability}, first studied by
Jona-Lasinio, Martinelli and Scoppola \cite{GMS}, and called by Simon
a ''flea on the elephant'' \cite{S4}, a phenomenon which has also dynamical
counterparts \cite{GSa}, \cite{WC}, and is at the root of the 
existence of the chiral superselection rules induced by the environment
which account for the shape of molecules such as ammonia, see the review
by A. S. Wightman \cite{Wightman}.

Unfortunately the s.c. spectrum does not posess either the dynamic or
the perturbation theoretic properties (see \cite{SWo}, \cite{Ho2}for
the latter) which are commonly associated with the physical picture of
delocalized states. Regarding the dynamical properties, for instance,
the \textbf{sojourn time} of a particle with initial state $\Psi$ in a
compact region  $S$ is defined as   
$$
J(S;\Psi) \equiv \int_{-\infty}^{\infty} ||P_{S} \exp(-itH) \Psi ||^{2} dt
\eqno{(5)}
$$
where $P_{S}$ denotes the projector on $S$. If $\Psi \in {\cal
  H}_{sc}$, it follows from a theorem by K.B. Sinha \cite{Si} that a
$S$ exists with $|S| < \infty$ and $J(S;\Psi) = \infty$. See also
\cite{MW}, \cite{MWB} for further discussion. 

We try therefore to attain higher dimensions.

\textbf{Multidimensional version}

Consider the Kronecker sum
$J_{\phi }^{\omega }$ as an operator on $\mathcal{H} \otimes \mathcal{H}$: 
\begin{eqnarray*}
J_{\theta }^{(2)}:=J_{\phi }^{\omega ^{1}}\otimes I+\theta I\otimes J_{\phi
}^{\omega ^{2}}  
\end{eqnarray*}
$$\eqno{(6)}$$ where $\omega ^{1}=\left( \omega _{j}^{1}\right)
_{j\geq 1}$ and $\omega 
^{2}=\left( \omega _{j}^{2}\right) _{j\geq 1}$ are two independent sequences
of independent random variables defined in $\left( \Omega ,\mathcal{B},\nu
\right) $, as before (we omit $\omega ^{1}$ and $\omega ^{2}$ in the l.h.s
above for brevity). Above, the parameter $\theta \in \lbrack 0,1]$
is included to avoid resonances . We ask for
properties of $J_{\theta }^{(2)}$ (e.g. the spectral type) which hold for 
\textbf{typical} configurations, i.e., a.e. $\left( \omega ^{1},\omega
^{2},\theta \right) $ with respect to $\nu \times \nu \times l$ where $l$ is
the Lebesgue measure in $[0,1]$. $J_{\theta }^{(2)}$ is a special
two--dimensional analog of $J_{\phi }^{\omega }$; if the latter was replaced
by $-\Delta +V$ on $L^{2}(\mathbb{R},dx)$ where $\Delta =d^{2}/dx^{2}$ is
the second derivative operator, and $V$ a multiplicative operator $V\psi
(x)=V(x)\psi (x)$ (potential), the sum above would correspond to $
\left( -d^{2}/dx_{1}^{2}+V_{1}\right) +\left( -d^{2}/dx_{2}^{2}+V_{2}\right) 
$ on $L^{2}(\mathbb{R}^{2},dx_{1}dx_{2})$, i.e., the
''separable case'' in two dimensions. Accordingly, we shall also refer to $
J_{\theta }^{(n)}$, $n=2,3,\ldots $, as the separable case in $n$ dimensions.

Our approach was to look at the quantity 

\begin{eqnarray*}
\left( \Phi ,e^{-itJ_{\theta }^{(2)}}\Psi \right) =f^{1}(t)f^{2}(\theta t)\\
\mbox{ where } \\
f^{i}(s)=f_{\mathrm{sc}}^{i}(s)+f_{\mathrm{pp}}^{i}(s)~,\quad i=1,2\\
\mbox{ with } 
f_{\mathrm{sc}}^{i}(s) &=&\int e^{-is\lambda }d\mu _{\varphi _{i},\psi
_{i}}^{\mathrm{sc}}(\lambda ) \\
f_{\mathrm{pp}}^{i}(s) &=&\int e^{-is\lambda }d\mu _{\rho _{i},\chi _{i}}^{
\mathrm{pp}}(\lambda )
\end{eqnarray*}

Above $\Phi ,\Psi \in \mathcal{H}\otimes \mathcal{H}$, 

\begin{eqnarray*}
\Phi &=&\left( \varphi _{1}\dot{+}\rho _{1}\right) \otimes \left( \varphi
_{2}\dot{+}\rho _{2}\right) ~,  \\
\Psi &=&\left( \psi _{1}\dot{+}\chi _{1}\right) \otimes \left( \psi _{2}\dot{
+}\chi _{2}\right) ~,
\end{eqnarray*}
with $\varphi _{i},\psi _{i}\in \mathcal{H}_{\mathrm{sc}}$, $\rho _{i},\chi
_{i}\in \mathcal{H}_{\mathrm{pp}}$ and $\varphi \dot{+}\rho $ denotes the
direct sum of two vectors $\varphi ,\rho \in \mathcal{H}$.

We shall use the following folklore proposition (see, e.g., theorem
3.2 of \cite{MWB} or \cite{Si}): 

\textbf{Proposition 1}

Let $\mu $ be a measure on the space $M(\mathbb{R})$ of all finite
regular Borel measures on $\mathbb{R}$. If the Fourier--Stieltjes transform
of $\mu $
\begin{eqnarray*}
\mathbb{R}\ni t\longmapsto \hat{\mu}(t)=\int e^{-it\lambda }d\mu (\lambda )
\end{eqnarray*}
belongs to $L^{2}(\mathbb{R},dt)$, then $\mu $ is absolutely continuous with
respect to Lebesgue measure.

The time-like decay of the Fourier-Stieltjes (F.S.) transform of the
spectral measure is dictated by the Hausdorff dimension of the
spectral measure: 

Let $S$ be a subset of $\mathbf{R}$, $\alpha \in [0,1]$, and $\delta > 0$. Define
\begin{eqnarray*}
h^{\alpha}_{\delta}(S) \equiv \inf \lbrace \sum_{j=1}^{\infty} |C_{j}|^{\alpha} \\
| S \subset \cup_{j=1}^{\infty} C_{j} \mbox{ with } |C_{j}| \le \delta \rbrace
\end{eqnarray*}
where $|C|$ denotes the Lebesgue measure (length) of $C$, and
$$
h^{\alpha}(S) = \lim_{\delta \to 0} h^{\alpha}_{\delta}(S) =
\sup_{\delta > 0} h^{\alpha}_{\delta}(S) 
$$
We call $h^{\alpha}$ $\alpha$ - dimensional Hausdorff measure on
$\mathbf{R}$. $h^{1}$ agrees with Lebesgue measure, and  
$h^{0}$ is the counting measure, so that $\{h^{\alpha} | 0 \le \alpha
\le 1\}$ is a family which interpolates continuously between the
counting measure and Lebesgue measure. 

\textbf{Definition 1} A Borel measure on $\mathbf{R}$ is
\textbf{uniformly alpha-Hoelder continuous} (U$\alpha$H) if there
exists $C$ such that for every interval $I$ with $|I| < 1$, 
$$
\mu(I) < C |I|^{\alpha}
$$

By an important theorem of Last \cite{La2}, U$\alpha$H measures may be
obtained by a process of closure. In our one-dimensional model the
Hausdorff dimension varies locally in the s.c. spectrum, and the local
Hausdorff dimension (suitably defined) may be determined
explicitly. There exists a dense set in the s.c. subspace such that
the spectral corresponding spectral measure is U$\alpha$H. We shall
for simplicity assume that the Hausdorff dimension has a constant
value which will be denoted by $\alpha$. 

The basic theorem, due to Strichartz \cite{Str} (with a very slick
alternative proof by Last \cite{La2}, therefore we call it
Strichartz-Last theorem), relates decay in the \textbf{Cesaro sense}
to the Hausdorff dimension: 

\textbf{Strichartz-Last theorem} 

Let $\mu$ be a finite U$\alpha$H measure and, for each 
$f \in L^{2}(\mathbf{R}, d\mu)$, denote
$$
\hat{f\mu}(t) \equiv \int \exp(-ixt) f(x) d\mu(x)
$$
Then there exists $C$ depending only on $\mu$ such that for all $f \in
L^{2}(\mathbf{R},d\mu)$ and $T > 0$, 
$$
\langle |\hat{f\mu}|^{2} \rangle_{T} < C ||f||^{2} T^{-\alpha}
$$
where $||f||$ denotes the $L^{2}$ - norm of $f$, and $\langle g
\rangle_{T} \equiv \frac{\int_{0}^{T}g(t)dt}{T}$. 

A measure is called a \textbf{Rajchman measure} iff
$$
\lim_{|t| \to \infty} \hat{\mu}(t) = 0
$$
It does \textbf{not} follow from the decay in the Strichartz-Last
theorem that the corresponding measure is Rajchman . Let $E$ denote
the usual Cantor ''middle-thirds'' set in $[-\pi,\pi]$; die
F.S. transform of the corresponding measure $\Gamma$ is: 
$$
\Gamma(u) = \prod_{j=1}^{\infty} \cos[2/3 u \pi 3^{-j+1}]
$$
The corresponding Hausdorff dimension is well-known to be
$\alpha= \frac{|\log 1/2|}{|\log 1/3|}$ und the measure is U$\alpha$H,
but, from the above formula for $\Gamma$ it follows that 
$$
\Gamma(n) = \Gamma(3n) \mbox{ for all } n \in \mathbf{Z}
$$
and hence the Cantor measure is not Rajchman.

The main result of \cite{MW} was:

\textbf{Theorem 2}
Let
\begin{eqnarray*}
v^{2}<a\left( \sqrt{\beta }-1\right) <v_{c}^{2}
\end{eqnarray*}
with $a<4$, where $v_{c} = 2\sqrt(\beta -1)$. Then, for almost every
$\left( \omega ^{1},\omega ^{2},\theta\right) $ with respect to $\nu
\times \nu \times l$, 
[a.] there exist $\tilde{\lambda}^{\pm }$ with $\tilde{\lambda}^{+}=-
\tilde{\lambda}^{-}$ and 
\begin{eqnarray*}
0<\tilde{\lambda}^{+}<\lambda ^{+} 
\end{eqnarray*}
such that
\begin{eqnarray*}
\left( \tilde{\lambda}^{-}(1+\theta ),\tilde{\lambda}^{+}(1+\theta )\right)
\subset \sigma _{\mathrm{ac}}\left( J_{\theta }^{(2)}\right)  
\end{eqnarray*}

[b.]
\begin{eqnarray*}
\left[-2(1+\theta ),\lambda ^{-}(1+\theta )\right) \cup \left( \lambda
^{+}(1+\theta ),2(1+\theta )\right] \subset \sigma _{\mathrm{pp}}\left(
J_{\theta }^{(2)}\right)
\end{eqnarray*}

[c.]
\begin{eqnarray*}
\sigma _{\mathrm{sc}}\left( J_{\theta }^{(2)}\right) \cap \left( \lambda
^{-}(1+\theta ),\lambda ^{+}(1+\theta )\right)  
\end{eqnarray*}
may, or may not, be an empty set.

The basic idea of the proof is that the Strichartz-Last theorem
suggests a pointwise decay of the F.S. transform of the spectral
measure of type $t^{-\alpha/2}$ for large $t$. For the Kronecker sum
(with $\theta=1$) the F.S. transform is the product of the
corresponding transforms for the one-dimensional system, which we
expect decays as $t^{-\alpha}$. In order to use the proposition, we
restrict the spectral measure to an interval in the s.c. spectrum such
that $2\alpha > 1$ which is in principle non-empty (and may be proven
so), and denote the F.S. transform of the thus restricted measure by
the same symbol as before. It turns, however, out that this heuristics
is not correct mathematically: as shown above, Cesaro decay
\textbf{does not} in general imply pointwise decay or, in other words,
that the spectral measure is Rajchman. This is a much harder problem -
(see \cite{MWB}, chapter 4 for the full treatment of the pointwise
decay of a model with superexponential sparsity and pure
s.c. spectrum, and the recent very hard analysis to prove local decay
in nonrelativistic QED \cite{CJFS}). Moreover, the result for general
$\theta$ may not be true due to resonance between Cantor sets, a
subtle phenomenon which has been so far analysed only in the
self-similar case by Hu and S.G. Taylor \cite{HuT}. Self-similarity
occurs, however, seldom: indeed, the s.c. spectrum of sparse Jacobi
matrices is not self-similar by a theorem of Combes and Mantica
\cite{CM}. We proved, however, that the main idea is indeed correct by
generalizing a method due to Kahane and Salem (\cite{KaS1},
\cite{KaS2}) in specific (self-similar) cases to the problem at hand.

We shall come back to this model in section 2d. 

\textbf{2b2 - Anderson model on the Bethe lattice}

Consider a particle moving - for simplicity - on the Bethe lattice, a  regular rooted tree graph,
with vertex set ${\cal G}$, in which each vertex, aside from the root $O$ has
$K+1$ neighbors with $K \ge 2$, and an edge set ${\cal E}$, described by a 
Schr\"{o}dinger operator of the form
$$
(H_{\lambda}(\omega) \Psi)(x) = -\sum_{(x,y) \in {\cal E}}\Psi(y) + \lambda V^{\omega}(x)\Psi(x)
\eqno{(7)}
$$
$V^{\omega}$ is a real valued random potential, with $\omega$ again representing the 
randomness. We assume, again for simplicity, that the $\{V^{\omega}(x)\}_{x \in {\cal G}}$ comprise
i.i.d. random variables, whose probability distribution is of the form $\rho(V)dV$ with $\rho$ the 
uniform distribution, constant over the interval $[-1,1]$ and zero elsewhere (the uniform distribution).
The regularity conditions assumed in \cite{AiWa3}, \cite{AiWa4}, \cite{AiWa5} are much more general,
but the uniform distribution is always a possible special case, and is the simplest example to be kept in mind.
$H_{\lambda}(\omega)$ is a bounded self-adjoint operator on $l^{2}({\cal G})$, with spectrum which is, for 
almost all realizations of $\omega$, the nonrandom set
$$
\sigma(H_{\lambda}) = [-|E_{\lambda}|,|E_{\lambda}|] \mbox{ with } E_{\lambda}=-(2\sqrt(K)+\lambda)
\eqno{(8)}
$$
The fact that the spectral measures associated to the different vectors $\delta_{x} \in l^{2}({cal G})$, where
$\delta_{x}$ denotes the Kronecker delta at $x$, are almost surely all equivalent, i.e., a.s. cyclicity of the
spectrum, is an important result of \cite{JaLa2}.

For a tree with $N$ generations, there are $N_{S}= (K+1) K^{N-1}$ vertices at the surface (generation $N = 0$)
and a total $N_{T} = (K^{N+1} + K^{N-2})/(K-1)$ vertices. Since the ratio $N_{S}/N_{T}$ is nonzero for
$N \to \infty$, the behavior of statistical systems on a Bethe lattice is quite distinct from those on a regular
Bravais lattice.

Models such as (7) have been considered since the early days of Anderson localization, see \cite{An1},\cite{Abou1},
\cite{Abou2}. The first mathematical proof of a transition was the seminal result of A. Klein \cite{Kl}:

\textbf{Theorem 3} (Theorem 1.1 of \cite{Kl}) For any $E$, $0<E<2\sqrt(K)$, there exists $\lambda(E) >0$ such
that for any $\lambda$ with $|\lambda| < \lambda(E)$ the spectrum of $H_{\lambda}$ in $[-E,E]$ is purely a.c.
with probability one. 

Aizenman \cite{Ai} proved localization in the Bethe lattice for energies beyond $K+1$, at weak disorder: this 
quantity appears as the edge of the $l^{1}$- spectrum of the free Laplacean on the lattice. What happens between
$2\sqrt(K)$ and $K+1$ for small $\lambda$? (This is problem 4 in Jitomirskaya's list of open problems). As noted
in \cite{J}, bottom of page 626, in analogy with the random surface case or the random decaying potentials 
mentioned in section 2a, one might conjecture that $l^{1}$- phenomena prevail up to the $l^{2}$- threshold, i.e.,
that there is localization up to near the edge of the $l^{2}$ spectrum - but see the following theorem 4b, due to
\cite{AiWa3}, \cite{AiWa4}, a surprising result which contradicted original expectations, and whose ideas we
now briefly sketch.

As explained in \cite{AiWa1}, \cite{AiWa3}, and stated heuristically, states which may locally appear to be
localized have arbitrarily close energy gaps $\Delta E$ with other states to which the tunneing amplitudes
decay exponentially in the distance $R$, as $\exp(-L_{\lambda}(E)R)$, where $L_{\lambda}(E)$ is the Lyapunov exponent
defined as follows. Let 
$$
G_{\lambda}(x,y;\zeta,\omega) = (\delta_{x}, (H_{\lambda}(\omega) - \zeta)^{-1}\delta_{y}))
$$
where $\zeta \in \mathbf{C}^{+} \equiv \zeta \in \mathbf{C} \mbox{ s.t. } \Im \zeta >0$ denote the Green's function.
Then
$$
L_{\lambda}(E) \equiv - \lim_{\eta \searrow 0}Av(|\log|G_{\lambda}(0,0;E+i\eta,\omega)|)
\eqno{(9-a)}
$$
where $Av$ stands for the expectation over the $\omega$ probability distribution. Indeed, it is shown in \cite{AiWa4}
that on tree graphs $L_{\lambda}$ provide the typical decay rate of the Green's function:
$$
\lim_{\eta \searrow 0}G_{\lambda}(0,x;E+i\eta,\omega) \sim C \exp(-L_{\lambda}(E)|x|)
\eqno{(9-b)}
$$
a.s. in $\omega$, where the limit exists for a.e. energy $E$ and $|x|=d(x,0)$ denotes the graph distance
of the vertex $x$ to some fixed but arbitrary vertex $O \in {\cal G}$. Mixing between two levels would occur if
$\Delta E \ll \exp(-L_{\lambda(E)}R)$, which is a very stringent condition, not satisfied in finite dimension, but,
since, as previously remarked, the volume grows exponentially, as $K^{R} = \exp(R\log K)$, extended states may
form, and a sufficient condition may be shown to be $L_{\lambda}(E) > \log K$:

\textbf{Theorem 4a} (Theorem 1 of \cite{AiWa3}, see also \cite{AiWa4} Under the previous assumptions, at a.e. 
$E \in \sigma(H_{\lambda})$ at which also 
$$
L_{\lambda}(E) > \log K
\eqno{(10-a)}
$$
one has, with probability one,
$$
\Im G_{\lambda}(x,x;E+i0,\omega) > 0
\eqno{(10-b)} 
$$
for all $x \in {\cal G}$.

Above, $E+i0$ denotes $\lim_{\eta \searrow 0}$ of the corresponding quantity with argument $E+i\eta$.
It is well known that
$$
\pi^{-1} \Im G_{\lambda}(x,x;E+i0,\omega)
\eqno{(10-c)}
$$
is the density of the a.c. component of the spectral measure associated with the state $|x> \equiv \delta_{x}$ (see,
e.g., \cite{Jak}, Theorem 3.15), and thus theorem 4a implies that in any energy interval within 
$\sigma(H_{\lambda}(\omega))$ for which (10a) holds, the random operator a.s. has some a.c. spectrum on a subset
of energies of positive Lebesgue measure. We now have:

\textbf{Theorem 4b} (Theorem 2 of \cite{AiWa3}) For
$$
\lambda < \Delta_{K} \equiv (\sqrt(K)-1)^{2}/2
\eqno{(10-d)}
$$
condition (10-a) holds for all energies satisfying
$$
|E| > |E_{\lambda}| - \delta(\lambda)
\eqno{(10-e)}
$$
at some $\delta(\lambda) > 0$.

Clearly (10-e) includes the spectral edges 
$[E_{\lambda},E_{\lambda}+\delta(\lambda)) \cup (|E_{\lambda}|-\delta(\lambda),|E_{\lambda}|]$. Thus, for $\lambda$ 
small enough (as in theorem 4b), near the spectrum's boundary the random operator has a.s. only purely a.c.
spectrum, and thus there is no mobility edge beyond which the spectrum is pure point and localization sets in -
a surprising result, which contradicts original expectations \cite{Abou1}\cite{Abou2}. On the basis of the joint
continuity of $L_{\lambda}(E)$ in $(\lambda,E)$, it has also been conjectured \cite{AiWa3} that for small 
enough $\lambda$ - for bounded potentials and sufficiently regular probability distributions - the spectrum is
purely a.c..

An improved version of theorem 4a includes information about large deviations, which are encoded in the
\textbf{free energy function}, defined for $s \in [-\kappa,1)$ - where $\kappa \in (0,1)$ is such that
$Av(V(0)^{\kappa}) < \infty$ - by
$$
\phi_{\lambda}(s;E) \equiv \lim_{|x| \to \infty}\frac{\log Av(|G_{\lambda}(0,x;E+i0,\omega)|^{s})}{|x|}
\eqno{(11-a)}
$$
and for $s = 1$ by $\phi_{\lambda}(1;E) \equiv \lim_{s \nearrow 1}\phi_{\lambda}(s;E)$. The existence of the limit
in (11-a), for a.e. $E \in \mathbf{R}$, is shown in (\cite{AiWa4}, Theorem 3.2). We have

\textbf{Theorem 4c} (Theorem 2.5 of \cite{AiWa4}) Under the previous assumptions (but, much more generally, 
assumptions A-E, pg.8, of \cite{AiWa4}), for any $\lambda > 0$ and Lebesgue a.e. $E \in \mathbf{R}$ at which
$$
\phi_{\lambda}(1;E) < -\log K
\eqno{(11-b)}
$$
the Green's function satisfies almost surely
$$
\Im G_{\lambda}(0,0;E+i0,\omega) > 0
\eqno{(11-c)}
$$

By convexity arguments (see, again, theorem 3.2 of \cite{AiWa4}), $\phi_{\lambda}(s;E) \ge -s L_{\lambda}(E)$, and,
hence, (11-b) implies (10-a) and, therefore, Theorem 4a is satisfied under the assumption of theorem 4a.

The importance of the free energy function is two-fold: it characterizes almost completely the different 
spectral regimes (see \cite{AiWa4}), and it is also a powerful tool in the \textbf{dynamics}, a topic to which 
we now turn our attention.

\textbf{2c - Dynamics}
 
We have seen the importance of the dynamics for spectral theory already in section 2a, through quantities
like the sojourn time (5) and the idea of the proof of theorem 2. This is also true regarding theorem 3, which
makes use of the fact that the FS transform of the spectral measure (associated to the Kronecker vector $\delta_{0}$
for instance), defined as in proposition 1, satisfies
\begin{eqnarray*}
\int_{-\infty }^{\infty }|\hat{\mu}(t)|^{2}dt\\
\leq \frac{1}{\pi }\liminf_{\eta \rightarrow 0+}\int_{-\infty }^{\infty }|G_{\lambda}(0,0;E+i\eta,\omega)|^{2}dE
\end{eqnarray*}
by the Plancherel theorem and Fatou's lemma (see \cite{Kl} or exercise 3.5 of \cite{MWB}). Klein succeeded in
proving that
$$
\liminf_{\eta \searrow 0}\int_{-E}^{E} |G_{\lambda}(x,x;E^{'}+i\eta,\omega)|^{2} dE^{'} < \infty
$$
with probability one for certain values of $(E,\lambda)$ mentioned in the statement of theorem 3, from which
the spectral measure restricted to $(-E,E)$ is purely a.c. by proposition 1.

Another interesting dynamical consequence of (11-c) was emphasized by Miller and Derrida \cite{MilD}, in a
basic theoretical paper on weak disorder expansions: if quantum particles are sent coherently at energy
$E= k^{2} + U_{\mbox{ wire }}$ down a wire attached to the graph at the vertex $x$, the reflection coefficient $R_{E}$
satisfies $R_{E} < 1$ iff (11-c) holds, see also the discussion in \cite{AiWa3}, pg.3.

A different class of problems involving the dynamics relates to \textbf{quantum dynamical stability}. We briefly
summarize the main issues, following (\cite{MWB}, section 3d). 
Most of the concepts of dynamical stability in quantum mechanics are related
to the growth of expectation values 
$$
\langle A(t)\rangle   
\eqno{(12-a)}
$$
of certain observables $A=A(0)$ as time $t$ evolves. \textbf{Boundedness} of 
$\langle A(t)\rangle $ is related to quantum stability. In general, for a lattice model
such as the the Anderson
model if the potential is random, or the case of an almost-periodic potential, it is
common to adopt $A=|X|^{2}$, where $X$ denotes the operator of multiplication by the
coordinate $X$, and one distinguishes the \textbf{localization regime}  
in which $\langle |X|^{2}\rangle <C$ for all $t$, 
or the \textbf{ballistic regime} where 
$$
\langle |X|^{2}\rangle =O(t^{2})\ ,\qquad 
\text{as }t\rightarrow \infty   
\eqno{(12-b)}
$$
which are supposed to correspond, respectively, to p.p. and a.c. spectrum;
Intermediate regimes are
characterized by the diffusion constant 
$$
D_{\Psi }=\lim_{t\rightarrow \infty }\frac{\langle |X|^{2}\rangle }{t}
\eqno{(12-c)}
$$
when the above limit exists , and the dependence on the initial state $\Psi $
is explicitly indicated. 

If $\langle |X|^{2}\rangle <C$ with $C$ a positive constant for all $t$,
then $\Psi $ has no continuous component. This is corollary 2.3.1. of 
\cite{La2}, and implies that $D_{\Psi }=0$. It is remarkable that the converse, i.e., that
if $\Psi \in \mathcal{H}_{\mathrm{pp}}$, then $D_{\Psi }=0$, is \textbf{not} true.
Indeed, the only general result about the converse is Simon's paper \cite{S2}
on the absence of ballistic motion, i.e., $\lim_{t\rightarrow
\infty }\langle |X|^{2}(t)\rangle =0$ if $\Psi \in \mathcal{H}_{pp}$, which
is far from the expected $D_{\Psi }=0$. The reason for this is the \textbf{
instability} of the two exotic spectra, dense p.p. and s.c.; indeed, even a
rank one perturbation with arbitrarily small norm is able to induce a
transition from one type to the other! (\cite{Ho1,SWo}). Since one
does not expect the dynamics to be strongly affected by such perturbations,
it is plausible that the absence of ballistic motion - the latter
characterizing a.c. spectra - is both a feature of s.c. spectra \textbf{and}
the (thick) p.p. spectra obtained from s.c. spectra by such perturbations:
therefore Simon's result might be optimal! This has indeed be shown in the
remarkable paper \cite{dR}, where a potential was constructed such that the
Hamiltonian $H$ on $l^{2}(\mathbb{Z})$ has a complete set of exponentially
decaying eigenfunctions but, for any $\delta >0$, $||X\exp (-itH)\delta
_{0}||^{2}/t^{2-\delta }$ is unbounded as $t\rightarrow \pm \infty $. Note
that $||x\exp (-itH)\delta _{0}||^{2}=(\exp (-itH)\delta _{0},X^{2}\exp
(-itH)\delta _{0})$. In the words of Jitomirskaya \cite{J}, this example
showed that mere \textquotedblleft exponential
localization\textquotedblright\ of eigenfunctions need not have any
consequence for the dynamics. Thus, \cite{dR} was pioneer in demonstrating
the importance of \textbf{dynamical localization} 
\index{dynamical localization}, which, since then, was proved for various
random models in the form 
\begin{eqnarray*}
\sup_{t}\left\Vert \left\vert X\right\vert ^{q}E_{I}(H_{\omega })\exp
(-itH_{\omega }\Psi )\right\Vert <\infty 
\end{eqnarray*}
with probability one. $I\subset \mathbb{R}$ is an energy interval in the
localization region and $E_{I}$ are spectral projections for $I$.
For the Anderson Hamiltonian, or the
deterministic almost-Mathieu model, several results are known,
and we refer to (\cite{J}, p. 621) for a discussion and (numerous)
references.

Concerning now the diffusion constant $D_{\Psi }$ given by (12-b),
one may define the so-called diffusion exponents (see, e.g.,
\cite{BS}):
\begin{eqnarray*}
\beta _{m}^{\pm }(\Psi )=\lim_{\pm}\frac{\log \langle |X|^{m}\rangle _{T}}{\log (T^{m})} 
\end{eqnarray*}
where $\lim_{\pm }\equiv \limsup_{T \to \infty}(\liminf_{T \to \infty} )$.
Both $\beta _{m}^{\pm }(\Psi )$ are
nonincreasing functions of $m$ and obey 
\begin{eqnarray*}
0\leq \beta _{m}^{-}(\Psi )\leq \beta _{m}^{+}(\Psi )\leq 1 
\end{eqnarray*}$$\eqno{(12-d)}$$

The following important theorem (\cite{Gu}, \cite{La2}, \cite{Co}) relates the
above quantities to the Hausdorff dimension of the s.c. spectrum:

\textbf{Guarneri-Last-Combes (GLC) theorem}
If $H$ is self-adjoint on $\mathcal{H}$ with spectral family $\{E(\lambda\}$
and uniformly alpha Hoelder continuous spectral measure $\mu_{\Psi}(\lambda)=(\Psi,E(\lambda)\Psi)$
(definition 1) there exists a constant $c_{\Psi ,m}$ depending on $\Psi $
and $m$ such that, for all $T>0$ and $m>0$, 
$$
\langle (\Psi (t),|X|^{m}\Psi (t))\rangle _{T}>c_{\Psi ,m}T^{m \alpha /d}
\eqno{(12-e)}
$$

Above, the time average $<.>_{T}$ is defined as in the statement of the Strichartz-Last theorem.
By the GLC theorem,
$$
\beta_{m}^{-}(\Psi) \ge \alpha/d
\eqno{(12-f)}
$$
For $\mu _{\Psi }$ a.c., $\alpha =1$; hence, in dimension 
$d=1$, a.c. spectrum implies ballistic transport, 
while in dimension $d=3$, a.c. spectrum and \textbf{subdiffusive motion}
($\beta _{m}<1/2$) may coexist. This motion is experimentally found in the so-called
\textbf{quasicrystals} (see \cite{GuSB} for a discussion and references). One of
the very few models in $d=3$ dimensions exhibiting a.c. spectrum with subdiffusive motion
this was constructed in \cite{BS} along lines similar
to those of theorem 2, but using one-dimensional Jacobi matrices with self-similar
s.c. spectra and making use of the aforementioned theorem of Hu and Taylor \cite{HuT}.

Defining the transition probability
$$
P_{\Psi,t}(x) \equiv |(\exp(-itH)\Psi)(x)|^{2}
\eqno{(13-a)}
$$
a different form of the time-average (for the observable $|x><x|$) is often easier to study \cite{AiWa5}:
\begin{eqnarray*}
\hat{P}_{\Psi,\eta}(x) \equiv 2\eta\int_{0}^{\infty} \exp(-2\eta t) P_{\Psi,t}(x) dt\\
= \frac{\eta}{\pi} \int |((H-E-i\eta)^{-1} \Psi)(x)|^{2} dE
\end{eqnarray*}$$\eqno{(13-b)}$$
with inverse time parameter $\eta > 0$ playing the role of $1/T$ in the aforementioned time-averages: the limit
of long time averages $T \to \infty$ becomes $\eta \searrow 0$. The equality (13-b) follows from the spectral
theorem and Plancherel's theorem and explains the utility of this type of averaging. The averages (12-a) may now be
written, with $A(0) = |X|^{\beta}$, 
$$
M_{\Psi}(\beta,t) = \sum_{x \in {\cal G}}|x|^{\beta} P_{\Psi,t}(x)
$$
and the modified time-averages as
$$
\hat{M}_{\Psi}(\beta,\eta) \equiv \sum_{x \in {\cal G}}|x|^{\beta} \hat{P}_{\Psi,\eta}(x)
$$
Ballistic (resp. diffusive) transport (12b)(resp. (12c)) correspond to $r=1$ (resp. $r= 1/2$) in the formulas:
$$
M_{\Psi}(\beta,t) \sim t^{r \beta}
\eqno{(14-a)}
$$
and
$$
\hat{M}_{\Psi}(\beta,\eta) \sim \eta^{-r \beta}
\eqno{(14-b)}
$$
in the finite-dimensional case ${\cal G} = \mathbf{Z}^{d}$.

A general ballistic upper bound
$$
\hat{M}_{\Psi}(\beta,\eta) \le C(\beta,\Psi) \eta^{-\beta}
\eqno{(14-c)}
$$
for all normalized $\Psi \in l^{2}({\cal G})$ with $\sum_{x \in {\cal G}}|x|^{\beta}|\Psi(x)|^{2} < \infty$, follows
from general arguments related to the Lieb-Robinson bounds \cite{NaSi}, see the appendix B of \cite{AiWa5}.

The strategy of proving the lower bound of the same type of the r.h.s. of (14-c) (with the purpose of showing
ballistic behavior) is of particular interest. By the GLC theorem, for ${\cal G} = \mathbf{Z}^{d}$, in the case
of uniformly alpha Hoelder continuous spectral measures, the Guarneri bound (see also \cite{Gu}) becomes
$$\hat{P}_{\Psi,\eta}(x) \le C \eta^{\alpha} \eqno{(15-a)}$$, and thus
\begin{eqnarray*}
\hat{Pr}_{\Psi,\eta}(|x|<b\eta^{-\alpha/d}) \equiv \sum_{|x|<b\eta^{-\alpha/d}}\hat{P}_{\Psi,\eta}(x,t)\\
< C_{d} b^{d}
\end{eqnarray*}
$$\eqno{(15-b)}$$
which implies that (14-a,b) can only hold if $r \ge \alpha/d$, because, selecting $b$ such that
$C_{d}b^{d}=1/2$ one obtains $\hat{M}_{\Psi}(\beta,\eta) \ge \frac{b^{\beta} \eta^{-\beta \alpha/d}}{2}$. 
This lower bound diminishes, however, as $d \to \infty$, and provides no information in this limit (which is
the tree graph case). A different method becomes, therefore, necessary, of which we sketch the main ideas.
With the notion of a.c. spectrum defined by
$$
\sigma_{ac}(H) \equiv \{E \in \mathbf{R} | Pr(\Im G(0,0;E+i0) \neq 0) > 0 \}
\eqno{(15-c)} 
$$
where $Pr$ denotes the probability associated to the random potential (we refer to \cite{AiWa5} for further
discussion of this concept, but remark that $\sigma_{ac}(H)$ is a nonrandom set which is a.s. the support of the
a.c. component of the spectrum), the following theorem holds:

\textbf{Theorem 5} (Theorem I.1 of \cite{AiWa5}) For any initial state of the form $\Psi = f(H) \delta_{0}$, with
a measurable function $f \in L^{2}(\mathbf{R})$ supported in $\sigma_{ac}(H)$, and all $b > 0$,
$$
Av(\hat{Pr}_{\Psi,\eta}(|x|<b\eta^{-1}) \le C(f) b + o(\eta)
\eqno{(15-d)}
$$
with some $C(f) < \infty$ and $o(\eta)$ a quantity which vanishes for $\eta \to 0$.

Theorem 5 yields a lower bound to $\hat{M}_{\Psi}(\beta,\eta)$ of the ballistic form r.h.s. of (14-c); together with
(14-c) it implies ballistic behavior for the full regime of the a.c. spectrum. Its significance is due to the
following facts: i.) in the hyperbolic geometry of a regular tree, the classical diffusion (which corresponds to
$r = 1/2$ in (14) in the finite dimensional case) also spreads ballistically, since, at each instant, there are
more directions at which $|x|$ would increase than the one direction at which it goes down ; ii.) it excludes the
possibility of a different behavior in the previously discussed regime where the a.c. spectrum is caused by rare
resonances, such as around the edges, where the density of states is very low and Lifschitz tail asymptotics
hold (see \cite{AiWa1}).

The following diffusive bound is (for a bounded measurable $I \subset \sigma_{ac}(H)$) the main ingredient of  
the proof of theorem 5:
$$
esssup_{\zeta \in I+i(0,1]} Av(|G(0,x;\zeta)|^{2}) \le \frac{C(I)}{K^{|x|}}
\eqno{(16-a)}
$$
where $C(I) < \infty$. The r.h.s. of (16a) coincides with the mean value of the total time spent at vertex
$x$ for a particle undergoing diffusion at the root $x = 0$. This bound is a consequence of the following bounds
for the free energy function (11a):
\begin{eqnarray*}
\phi(s;\zeta) \le -s \log K \mbox{ for all } \zeta \in \mathbf{C}^{+}\\
\mbox{ and for all } s \in [0,2]
\end{eqnarray*}
$$\eqno{(16-b)}$$
and
\begin{eqnarray*}
Av(|G(0,x;\zeta)|^{s} \le C_{+}(s,\zeta) \exp[\phi(s;\zeta)|x|] \\
\mbox{ for all } s \in [0,1) \mbox{ and } x\in {\cal G}
\end{eqnarray*}$$\eqno{(16-c)}$$
where $C_{+}(s,\zeta) \in (0,\infty)$ uniformly in $\zeta \in [-E,E]+i(0,1]$ for any $0 \le E < \infty$
at a fixed $s \in [0,1)$.

Since for energies in the p.p. spectrum $Av(|G(0,x;E+i0)|) = \infty$ (see, e.g., \cite{Jak}, theorem 3.15), (16c)
cannot be expected to hold beyond $s = 1$, but theorem II.2 of \cite{AiWa5} shows that this bound extends to
all energies such that
$$
esssup_{\zeta \in I+i90,1]} Av((\Im G(0,0;\zeta))^{-3-\delta}) < \infty
\eqno{(16-d)}
$$
for some $\delta > 0$, and then theorem II.4 of \cite{AiWa5} shows that (16-d) includes any bounded measurable
set $I \subset \sigma_{ac}(H)$, with $\sigma_{ac}(H)$ given by (15-c).

\textbf{2d - Concluding remarks on the Anderson transition}

An essential point, noted in \cite{J}, is that Klein's proof of theorem 3 uses the looples character of the graph
in an essential way and, thus, the Bethe lattice, while corresponding to the limit $d \to \infty$, is, in a
sense, one-dimensional. Similarly, the sparse model of section 2b1 also has no loops, because theorem 2 ''inherits''
the one-dimensional structure of the model of theorem 1. Thus, both models still run short of solving the central
issues of the original model, which remains one of the basic challenges of mathematical physics.

The model on the Bethe lattice describes the limit $d \to \infty$ and, as remarked in \cite{AiWa3}, the new and
surprising phenomena found there may even have wider applicability due to the analogies drawn between tree graphs
and many-particle configurations \cite{Alt}. They are, however, not expected in finite dimensions, in particular the
absence of the mobility edge for weak disorder (Theorem 4b) is not found in the sparse model
(see theorem 1 and theorem 2, where the existence of p.p. spectrum in a nonempty energy interval for arbitrarily small
disorder is pictured as a concrete example of the tunneling instability). However, the methods used to prove
theorem 4b, in particular concerning the free energy function (11a), may have wider applicability, in particular
to investigate the existence or not of a sharp mobility edge in other models, such as the one in section 2b1 or
Krishna's model \cite{Kri3}. In addition, the ideas developed for the Bethe lattice dynamics reviewed in 2c, such
as related to the negative moments of the Green's function (16-d), also deserve special attention as potentially
powerful methods to prove existence of the a.c. spectrum.

On the other hand, sparse models such as the one depicted in section 2b1, in dimension $d \ge 2$, comprise a part
of the full model on a Bravais lattice. Although exponential sparsity is too severe, it should be recalled 
that the separable model (6) does not take account of dimensionality in a proper way, because in ''truly d-dimensional''
sparse models the cardinality of the set in (4) may change from $O(\log R)$ to $O(R^{d-\epsilon})$, for some 
$\epsilon > 0$, and ''filling in'' the remaining points may be expected to remedy this flaw, but this is an immensely
challenging open problem! The sparse version has, however, one promising feature as compared with the full Anderson
model: in the latter, the $d \ge 2$ version built as in (6) from the one-dimensional version (the separable full
Anderson model) continues to have purely p.p. spectrum, by \cite{GMP} and the fact that convolutions of p.p.
measures are p.p., in complete disagreement with the expected transition. It may thus seem worthwhile to study
sparse models such as that of 2b1 as prospective descriptions of lightly doped semiconductors, which are also
expected to exhibit an Anderson transition in dimension $d \ge 2$ \cite{ShEf}. Of course, non-sparse models such as
the one proposed in \cite{Kri3}, are also very promising, although the available picture of the transition seems
to be, as yet, somewhat less detailed than the model described in section 2b1. 

\textbf{3-Frustration and short-range spin glasses: the Edwards-Anderson model}

Dilute solutions of atoms of large magnetic moments (such as the
transition metals Fe, Co, Mn) in a paramagnetic substrate (Cu, Au)
present a number of peculiar physical properties. For small, but
sufficiently high concentrations of magnetic impurities, the
susceptibility in low fields displays a characteristic peak, with
discontinuous derivatives, at a temperature $T_{g}$. The specific heat
is always smooth, with a linear dependence on the temperature as $T
\to 0$. The behaviour of these \textbf{spin glasses} has been
explained in terms of an indirect RKKY interaction between the spins
of these magnetic impurities mediated by the electrons of the
paramegnetic matrix \cite{BiYo}, which is of the form
$-J(|i-j|)S_{i}S_{j}$ with  
$J(r) = (k_{F}r)^{-3}\cos(2k_{F}r)$, where $r$ is the distance between
magnetic atoms and $k_{F}$ the fermi momentum;  
$S_{i}=\pm 1$ are (e.g.) Ising spins. The rapid oscillations and the
weak decay of $J(r)$, as well as the random distribution of the
impurity magnetic atoms, are the basic ingredients of spin glasses. At
sufficiently low temperatures there is a ''freezing'' of the magnetic
moments in random directions (which leads to an increase of the
susceptibility). The spin glass phase may be regarded as a
conglomerate of blocks of spins, each block with its own
characteristic orientation, in such a way that there is no macroscopic
magnetic moment \cite{BiYo}. 

Edwards and Anderson \cite{EA} proposed a spin Hamiltonian, which we
write in a generalized version as follows. 
For each finite set of points $\Lambda \subset \mathbf{Z}^{d}$, where
the dimension $d$ will be restricted to the values $d=2$ and $d=3$,
consider the Hamiltonian 
$$
H_{\Lambda}(\{J\}) = \sum_{i,j \in \Lambda} J_{i,j} \Phi_{i,j}
\eqno{(17-a)}
$$
where $\Phi_{i,j}$ with $i,j \in \Lambda$ are self-adjoint elements of
the algebra generated by the set of spin operators, the Pauli matrices
$\sigma_{i}^{x}$, $\sigma_{i}^{y}$, $\sigma_{i}^{z}$, $i \in \Lambda$,
on the Hilbert space ${\cal H}_{\Lambda} = \otimes_{i \in \Lambda}
\mathbf{C}_{i}^{2}$, given by 
$$
\Phi_{i,j} = \alpha_{x} \sigma_{i}^{x}\sigma_{j}^{x} + \alpha_{y}
\sigma_{i}^{y}\sigma_{j}^{y}  
+\alpha_{z}\sigma_{i}^{z}\sigma_{j}^{z}
\eqno{(17-b)}
$$
for $|i - j| = 1$, and zero otherwise. The random couplings $J_{i,j}$,
with $|i - j| = 1$, are random variables (r.v.) on a probability space
$(\Omega, {\cal F}, P)$, where ${\cal F}$ is a sigma algebra on
$\Omega$ and $P$ is a probability measure on $(\Omega,{\cal F})$. We
may take without loss of generality  
$$\Omega = \times_{B^{d}} S \eqno{(18)}$$ where $S$ is a Borel subset
of $\mathbf{R}$, $B^{d}$ is the set of bonds in $d$ dimensions, and
assume that the $J_{i,j}$ are independent, identically distributed
r.v.. In this case, $P$ is the product measure $$dP = \times_{B^{d}}
dP_{0} \eqno{(19)}$$ of the common distribution $P_{0}$ of the random
variables, which will be denoted collectively by $J$. The
corresponding expectation (integral with respect to $P$) will be
denoted by the symbol $Av$. We have to assume that $$ Av(J_{i,j}) = 0
\eqno{(20)}$$ for all $i,j \in \mathbf{Z}^{d}$, i.e., that the
couplings are centered. This assumption mimicks the rapid oscillations
of the RKKY interaction. Let $E_{\Lambda}$ denote the GS energy of
$H_{\Lambda}$, i.e., $E_{\Lambda} = \inf spec(H_{\Lambda})$. The
following result was proved, among several others, in \cite{ConL}: 

\textbf{Theorem 6} (\cite{ConL}) For $P$- a.e. $\{J\}$, the limit
below exists and is independent of the b.c.: 
$$
e^{(d)} \equiv \lim_{\Lambda \nearrow \mathbf{Z}^{d}} \frac{E_{\Lambda}}{|\Lambda|}
\eqno{(21-a)}
$$
and
$$
e^{(d)} = \inf_{\Lambda} \frac{E_{\Lambda}}{|\Lambda|}
\eqno{(21-b)}
$$
where $|\Lambda|$ denotes the number of sites in $\Lambda$. Finally,
$$
e^{(d)} \ge e^{(d+1)}
\eqno{(21-c)}
$$

(21-a) is the far-reaching property of self-averaging (see \cite{An3}
for a discussion): it expresses that measurable - e.g. thermodynamic-
quantities are the same for any \textbf{typical} configuration of the
sample, i.e., are experimentally reproducible. It follows from (21a)
that, $P$- a.e., 
$$
e^{(d)} = \lim_{\Lambda \nearrow \mathbf{Z}^{d}} \frac{Av(E_{\Lambda})}{|\Lambda|}
\eqno{(21-d)}
$$
Let $\Lambda_{N}$ denote a square with $N$ sites if $d=2$ or a cube
with $N$ sites if $d = 3$, and write  
$H_{N}^{(d)}(J) \equiv H_{\Lambda_{N}}(J)$. We now adopt periodic
b.c. for simplicity, but the final result is independent of the
b.c. due to theorem 1. We may write 
$$
H_{N}^{(d)}(J) = \sum_{n \in \Lambda_{N}} H_{n}^{(d)}(J)
\eqno{(22-a)}
$$
where $H_{n}^{(d)}$ is given by
$$
H_{n}^{(d)}(J) = c_{d} \sum_{(i,j) \in \Lambda_{n}^{(d)}} J_{i,j}\Phi_{i,j} 
\eqno{(22-b)}
$$
Above, $c_{d}$ are factors which eliminate the multiple counting of bonds, i.e.,
$$
c_{2} = 1/2
\eqno{(22-c)}
$$
and
$$
c_{3} = 1/4
\eqno{(22-d)}
$$
and $\Lambda_{n}^{(2)}$ is a square labelled by a site $n$, for which
we adopt the convention, using a right-handed $(x,y)$ coordinate
system, that $n=(n_{x},n_{y})$ is the vertex in the square with the
smallest values of $n_{x}$ and $n_{y}$. Similarly, $\Lambda_{n}^{(3)}$
is a cube labelled by a site $n = (n_{x},n_{y},n_{z})$ with, by the
same convention, the smallest values of $n_{x}$, $n_{y}$ and
$n_{z}$. Due to the periodic b.c., the sum in (22-b) contains
precisely $N$ lattice sites. The notation $H_{n}^{(d)}$ is short-hand
for its tensor product with the identity at the complementary lattice
sites in  
$\Lambda_{N}\backslash \Lambda_{n}^{(d)}$.
Let $E_{0}^{(N,d)}(J)$ denote the GS energy of $H_{N}^{(d)}(J)$ ,
$E_{0}^{(n,d)}(J)$ the GS energy of $H_{n}^{(d)}(J)$ and
$$
E_{0}^{(d)} \equiv Av(E_{0}^{(n,d)}(J))
\eqno{(22-e)}
$$
By the condition of identical distribution of the r.v. $J_{i,j}$,
$E_{0}^{(d)}$ does not depend on $n$, which is implicit in the
notation used. We have 

\textbf{Theorem 7} (Theorem 2 and proposition 1 of \cite{Wr5}) The
following lower bound holds: 
$$
\frac{Av(E_{0}^{(N,d)}(J))}{N} \ge E_{0}^{(d)}
\eqno{(23-a)}
$$
Further, let $dP_{0}$ in (9) be the Bernoulli distribution $dP_{0} =
1/2(\delta_{J} + \delta_{-J})$, set for simplicity $J = 1$ and let
$\alpha_{x} = \alpha_{y} = 0$ in (17b). Then, 
$$
e^{(2)} \ge -3/2
\eqno{(23-b)}
$$
and
$$
e^{(3)} \ge -1/4\frac{36096}{4096}
\eqno{(23-c)}
$$

The special case $\alpha_{x} = \alpha_{y} = 0$ in (17-b) is the
classical EA spin-glass; we also set $\alpha_{z} = 1$. In this case
$e^{(d)}$, given by the r.h.s. of (21-d), is invariant under the
''local gauge transformation''  
$\sigma_{i}^{z} \to -\sigma_{i}^{z}$ together with $J_{i,j} \to
-J_{i,j} \mbox{ for all } j | |j-i|=1$, whatever the lattice site $i$.  

According to the above, \cite{To}, an elementary square
(''plaquette'') $P$ is said to be \textbf{frustrated}
(resp. non-frustrated) if $G_{P} \equiv \prod_{(i,j) \in P} J_{i,j} =
-1$ resp. $+1$. Note that $G_{P}$ is gauge-invariant and that, for the
quantum XY (or XZ) model defined by setting $\alpha_{y} = 0$ in (17-b),
$e^{(d)}$ is also locally gauge-invariant if we add to the above
definition the transformation $\sigma_{i}^{x} \to -\sigma_{i}^{x}$,
i.e., the transformation on the spins is defined to be a rotation of
$\pi$ around the $y$ - axis in spin space. The property of local
gauge-invariance guarantees the absence of a macroscopic magnetic
moment (spontaneous magnetization) mentioned before as a basic
property of spin glasses \cite{ARS}. 

Since the Pauli z-matrices commute, finding the minimum eigenvalue of
(17-a) in the classical EA case is equivalent to find the configuration
of Ising spins $\sigma_{i} = \pm 1$, denoted collectively by $\sigma$,
which minimizes the functional 
$$
F(\sigma,J) \equiv \sum_{\sigma} J_{i,j} \sigma_{i} \sigma_{j}
$$ 
The minimal energy of a frustrated plaquette equals $E_{P,f} = -2$ and
of a non-frustrated plaquette $E_{P,u} = -4$.  

(23-b,c) of theorem 7 provide the first (nontrivial) rigorous lower
bound both for $e^{(2)}$ and $e^{(3)}$. Using the natural misfit
parameter $$m = \frac{|E^{id}| - |E_{0}|}{|E^{id}|} \eqno{(24)}$$ as a
measure of plaquettes frustrated or bonds unsatisfied (see (4) of
\cite{KK}), where $E_{0}$ denotes the ground state energy of the
frustrated system and $E^{id}$ is the ground state energy of a
relevant unfrustrated reference system, we find from (23-b) in the $d =
2$ case the lower bound $m \ge 0.25$ and for $d = 3$, from (23-c), the
lower bound $m \ge 0.26\cdots$: thus, in both cases, the measure of
frustrated plaquettes or unsatisfied bonds as defined above is at
least of the order of $1/4$. 

The method of proof of theorem 7 is a rigorous version of a
finite-size cluster method, originally due to \cite{BaS}, together
with the variational principle. If we take in (17) $\alpha_{z} = 1$,
$\alpha_{y} = 0$, and $\alpha_{x} = \alpha$, and consider $\alpha$ as
a small parameter, we have the anisotropic XZ (or XY) model. By the
norm-equicontinuity (in the volume) of $H_{\Lambda}(\{J\})$ (given by
(17)) as a function of $\alpha$, which is preserved upon taking
averages over the probability distribution of the $J$, it follows that
(23-b,c) hold with the right hand sides varying by small amounts if
$|\alpha|$ is sufficiently small: this is conceptually important for
reasons mentioned in the introduction. 

The mean field theory, recently rigorously solved 
(see \cite{Gue} for a review) does not exhibit frustration - indeed,
this concept is not even generally defined for their model, since the
theory does not require an underlying lattice. Whether frustration is
an important issue in the description of realistic spin glasses is an
important open problem. 

We refer to the article by Bovier and Fr\"{o}hlich \cite{BoFr} (see
also Bovier's book \cite{Bo}) for an illuminating discussion of
complementary, mostly global (i.e., involving the lattice as a whole)
aspects of frustration. Our bound (23-c), which relies only on the
\textbf{local} structure, is, however, slightly better than
Kirkpatrick's \cite{SKir}, which is based on very reasonable, but
unproved conjectures of a global nature. A most relevant open problem
would be, of course, to extend the finite-size cluster method to
obtain bounds for the free energy of the EA model. 

\textbf{4- Return to equilibrium and (thermodynamic) instability of the first and second kinds}

The dynamics of spin glasses is a topic of great relevance, both
conceptual and experimental. It happens, however, that the standard
approach to the kinetic theory (e.g., for the mean-field spin glass
model) relies on Glauber dynamics (see, e.g., \cite{Sz}),a well-known
dynamics imposed on the Ising model, which has been used to study
metastability in early days \cite{CCO} and more recently
\cite{SchS}. In spite of the considerable independent mathematical
interest of the latter works, it should be remarked that only for a
quantum system does a physically satisfactory definition of the
dynamics of the states and observables exist which is relevant to the
microscopic domain, including, of course, condensed matter
physics. The scarcity of examples of approach to equilibrium in
quantum mechanics is due, of course, to the extreme difficulty of
estimating  specific properties of the quantum evolution.  

This fact was the basic motivation which prompted Emch \cite{Em} and
Radin \cite{Ra} to propose a quantum dynamical model (which we dubbed
the Emch-Radin model in \cite{Wr5} and \cite{MWB}), which is relevant
to systems with high anisotropy, and displays a remarkable property of
\textbf{non-Markovian approach to equilibrium}, or return to
equilibrium. It turns out that such a model is useful not only for a
description of nuclear spin-resonance experiments - such as the one
\cite{LNo} which motivated Emch and is described below - but also to
describe dynamical effects associated to quantum crossover phenomena
in spin glasses (see \cite{Sach} for a review and references). For
this purpose, a material was chosen with a strong spin-orbit coupling
between the spins (of the magnetic ion) and the underlying crystal:
this coupling essentially restricts the spins to orient either
parallel or antiparallel to a specific crystalline axis, which we
shall label as the z-axis. Such spins are usually referred to as Ising
spins, and will be described by the Ising part of the forthcoming
Hamiltonian. In the experiments (see \cite{Sach} and references), a
transverse magnetic field is then applied, oriented perpendicular to
the z axis. A large enough transverse field will eventually destroy
the spin glass order even at $T=0$, leding to the existence of a
crossover region. 

We shall be interested in a situation in which a large transverse
field has been applied; the \textbf{initial} state of the system is,
then, approximately, a product state of the forthcoming form (16). The
same experimental setup should therefore allow a measurement of the
rate of \textbf{return to equilibrium} of the mean transverse
magnetization, i.e., of how $f(t)$, defined by (20), tends to zero, or
of how (21) is approached. As we shall see, this rate depends strongly
on the probability distribution of the $J$.    

The XY model has also been extensively studied from the point of view
of return to equilibrium, since early days as an example of general
theory \cite{Ro}, more recently in \cite{AB} (see the latter for
reference to related important work of Araki), but unfortunately only
in one dimension.

\textbf{The nonrandom model}

Following Radin's review \cite{Ra} of the work of Emch \cite{Em},
consider the experiment of \cite{LNo}: a $CaF_2$ crystal is placed in
a magnetic field thus determinig the z-direction, and allowed to reach
thermal equilibrium. A rf pulse is then applied which turns the net
nuclear magnetization to the $x$ direction. The magnetization in the
$x$ direction is then measured as a function of time.  

As in (\cite{Em},\cite{Ra}), we assume an interaction of the form
$$
H_{V}=1/2\sum_{j,k}\epsilon(\vert j-k \vert)\sigma^{z}_{j}\sigma^{z}_{k} 
\eqno{(25)}
$$
where $j\in V$ and $k\in V$ in (15), with $V$ is a finite region,
$V\in\mathbf{Z}^{d}$. $H_{V}$ is defined on the Hilbert space ${\cal
  H}=\bigotimes_{i\in V} \mathbf{C}^{2}$. The state representing the
system after the application of the rf pulse will be assumed to be the
product state 
$$
\rho=\rho(0)=\bigotimes_{j\in V}\phi_{j}
\eqno{(26-a)}
$$

where

$$
\phi_{j}(\cdot)=tr_{j}(\cdot \exp(-\gamma\sigma^{x}_{j}))/tr
_{j}(\exp(-\gamma\sigma^{x}_{j}) 
\eqno{(26-b)}
$$
and $tr$ is the trace on $\mathbf{C}^{2}_{j}$. Other choices of the
state are possible \cite{Ra}, but we shall adopt (16) for
definiteness. Let 
$$
S^{x}=1/N(V)\sum_{j\in V}\sigma^{x}_{j}
\eqno{(27)}
$$
be the mean transverse magnetization, with $N(V)$ denoting the number
of sites in $V$. The real number $\gamma$ in may be chosen as in
\cite{Em} such as to maximize the microscopic entropy subject to the
constraint $tr S^{x}\rho(0)$ equal to a constant, i.e., a given value
of the mean transverse magnetization. Since the state (26) is a
product state, 
$\gamma$ is independent of $V$, and has the same value if $S^{x}$ is
replaced by $\sigma^{x}_{i_0}$, for any $i_0\in V$. Let 
$$
\rho^{V}_{t}\equiv U_{V}^{t} \rho(0)U_{V}^{-t}
\eqno{(28-a)}
$$
and
$$
\langle\sigma^{x}_{i_0}\rangle_{V}(t)\equiv \rho^{V}_{t}(\sigma^{x}_{i_0})
\eqno{(28-b)}
$$
where $U_{V}(t)=\exp(itH_{V})$. We may write, by (18),
$$
\langle\sigma^{x}_{i_0}\rangle_{V}(t)=\rho(0)(U_{V}(-t)\sigma^{x}_{i_0}U_{V}(t))
\eqno{(28-c)}
$$
It is natural to define
$$
\langle\sigma^{x}_{i_0}\rangle(t)=\lim_{V\to\infty}\langle\sigma^{x}_{i_0}\rangle_{V}
(t)     
\eqno{(29)}
$$
provided the limit on the r.h.s. of (29) exists, as the expectation
value of the local transverse spin in the (time-dependent)
nonequilibrium state $\rho_{\infty}^{t}$ of the infinite system. A
weak-star limit of the sequence of states  
$\rho_{V}^{t}(\cdot)= tr(\cdot U_{V}^{t}\rho(0)U_{V}^{-t})$ exists, by
compactness, on the usual quasilocal algebra  
$\cal{A}$  of observables (see, e.g. \cite{Se3}, for a comprehensive
introduction to the algebraic concepts employed here). The expectation
value of $\sigma^{x}_{i_0}$ in the equilibrium state associated to
(25) is zero by the symmetry of rotation by $\pi$ around the $z$
axis. We may now pose the question whether the limit 
$$
f(t)\equiv
\lim_{N(\Lambda)\to\infty}\rho_{\infty}^{t}(1/N(\Lambda)\sum_{i_0\in\Lambda}
\sigma^{x}_{i_0})   
\eqno{(30)}
$$
where $\Lambda$ denotes a finite subset of $\mathbf{Z}^{\nu}$, which
is interpreted as the mean transverse magnetization, exists. The
property of return to equilibrium is expressed  by  
$$
\lim_{t\to\infty}f(t)=0
\eqno{(31)}
$$  
Of particular interest is the rate of return to equilibrium. In the
present nonrandom case, the limit at the r.h.s. of (31) equals
$f(t)=\rho_{\infty}^{t}(\sigma^{x}_{i_0})$, for any $i_0$, by
translation invariance of $\rho_{\infty}^{t}$. This is not so for
random systems, in which case additional arguments are necessary to
show the convergence of the r.h.s. of (31) for almost all
configurations of couplings. 

We assume that the function $\epsilon(\cdot)$ satisfies:
$$
\epsilon(0)=0
\eqno{(32)}
$$
and either:
$$
\sum_{j \in \mathbf{Z}^{d}}  \epsilon(j)  < \infty \mbox{ i.e. }
\epsilon\in l_1(\mathbf{Z}^{d}) 
\eqno{(33-a)}
$$
or
$$
\sum_{j \in \mathbf{Z}^{d}} (\epsilon(j))^{2} < \infty  \mbox{ i.e. }
\epsilon\in l_2(\mathbf{Z}^{d}) 
\eqno{(33-b)}
$$

\textbf{ The random model}

We now introduce the Hamiltonian of a disordered system corresponding
to (25), where we take, for simplicity,  
$$V = V_{n} = [-n,n]^{d} \eqno{(33-d)}$$ :
$$
\tilde{H_{n}}=1/2\sum_{j,k \in V_{n}} J_{(j,k)}\epsilon(j-k)\sigma^{z}_{j}\sigma^{z}_{k}
\eqno{(34)}
$$
where $J_{(j,k)}$ are independent, identically distributed random
variables (i.i.d. r.v.) on a probability space which we denote by
$(\Omega,{\cal B}, P)$. We shall use  $Av(\cdot)$ to denote averaging
with respect to the random configuration $\{J_{j,k}\}$, which we
denote collectively by the symbol $J$. The $J$ are assumed to satisfy
\cite{vEvH}: 
$$
Av(J_{(j,k)})=0
\eqno{(35-a)}
$$
$$
| Av(J^{n}_{(j,k)})| \le n!c^{n} \forall n=2,3,4,\cdots
\eqno{(35-b)}
$$
We shall take, without loss, $\epsilon(j) \ge 0$. If
\begin{eqnarray*}
\epsilon(j) = \left\{
\begin{array}{ll}
\beta & \mbox{ if } j \in \pm \delta_{i}\,,\\
0     & \mbox{ otherwise }\,,
\end{array}
\right.
\end{eqnarray*}
$$\eqno{(35-c)}$$
where $\pm \delta_{i}$ for $i=1, \cdots, d$ denote the set of $z=2d$
bonds connecting the origin to a point of $\mathbf{Z}^{d}$, we have
the EA spin glass model of section 3.  

Let $|V_{n}|$ denote the number of sites in $V_{n}$ and the free
energy per site $f_{n}$ be defined by 
$$
f_{n}(J)\equiv \frac{-kT\log Z_{n}(J)}{|V_{n}|}
\eqno{(36-a)}
$$
where
$$
Z_{n}(J)= tr(\exp(-\beta\tilde{H_{n}}))
\eqno{(36-b)}
$$
is the partition function and the trace is over the Hilbert space
${\cal H}$. Then 

\textbf{Theorem 8}\cite{vEvH} (see also \cite{KhSi} for the first
result in this direction):

Under assumptions (22-b) and (35), the thermodynamic limit of the free
energy per site 
$$
f(J)=\lim_{n\to\infty} f_{n}(J)
\eqno{(37-a)}
$$
exists and equals its average:
$$
f(J)= Av(f(J))=\lim_{n\to\infty} Av(f_{n}(J))
\eqno{(37-b)}
$$
for almost all configurations $J$ ($\mbox{ a.e. } J $).

The reason why (33-b) suffices for the existence of the thermodynamic
limit is that, in order to obtain a uniform lower bound for the
average free energy per site, the cumulant expansion (see, e.g.,
(\cite{S6}, (12.14), pg 129, for the definition of $Av_c$, there
called Ursell functions): 
\begin{eqnarray*}
Av(\exp(tJ_{(i,j)})= \exp(\sum_{n=2}^{\infty}Av_c(J^{n}_{(i,j)})t^{n}/n!)
\end{eqnarray*}
was used \cite{vEvH}, which, by (35-a), starts with the
\textbf{second} cu\-mu\-lant \hfill \break $Av_c(J^{2}_{(i,j)})=Av(J^{2}_{(i,j)})$,
which is the variance of $J_{(i,j)}$. Condition (35-b) was used to
control the sum in the exponent above.  
 
\textbf{Thermodynamic stability}

Stability considerations play a key role in quantum mechanics
\cite{LiS}. Let a statistical mechanical system be described by a
collection of amiltonians $H_{\Lambda}$, associated to finite regions
$\Lambda \subset \mathbf{R}^{d}$, for particle systems, or $\Lambda
\subset \mathbf{Z}^{d}$ for spin systems, with volume $V(\Lambda)$:
examples are (25) and, for random systems,(34): in the latter case
there is implicit in the Hamiltonians a dependence on the random
variables $J$, and the constant $c$ in definition 2 below is assumed
to be a.e. independent of $J$. The system's free energy is 
$f_{\Lambda}\equiv \frac{-kT\log Z_{\Lambda}(J)}{V(\Lambda)}$ with
$Z_{\Lambda}= tr(\exp(-\beta\ H_{\Lambda}))$, and the thermodynamic
limit means $\Lambda \nearrow \mathbf{R}^{d}$ (or $\mathbf{Z}^{d}$)
with the proviso of fixed density for particle systems, where $\Lambda
\nearrow \mathbf{Z}^{d}$ denotes a limit in the sense of van Hove or
Fisher (the latter being required for random systems, see the
discussion in \cite{vEvH}: (33-d) satisfies this assumption). 

\textbf{Definition 2} 
The system is said to be \textbf{thermodynamically stable} if there
exists a constant  
$0 \le c < \infty$ such that
$$
f_{\Lambda} \ge -c
\eqno{(38-a)}
$$
It is said to satisfy \textbf{thermodynamic stability of the first kind} if
$$
H_{\Lambda} > - \infty
\eqno{(38-b)}
$$
and to satisfy \textbf{thermodynamic stability of the second kind} if
$$
H_{\Lambda} \ge -c V(\Lambda)
\eqno{(38-c)}
$$
The above definitions (38-b,c) are patterned after the corresponding
ones for $N$- body systems in \cite{LiS}: they also apply to
relativistic quantum field theory, where the particle number $N$ is
not conserved. 

Theorem 5 illustrates the interesting fact that, for disordered
systems, stability of the second kind (38-c) may fail even for a
thermodynamically stable system: this happens when only (33-b) (but
not (33-a)) holds. In the following, we shall demonstrate that this
phenomenon has important \textbf{dynamical} implications for a
relevant class of disordered systems. 

We now return to our nonrandom model (25). When (33-a) holds, it
follows from a folklore theorem (see, e.g., Theorem 6-1 of \cite{MWB}
or Theorem 3.3, pg. 111, of \cite{Dav}) and a representation of $f$
(\cite{Ra}, pg.295, and proposition 1) that exponential decay in the
sense that 
$$
|f(t)| \le C \exp(-d|t|) 
\eqno{(39)}
$$
with $C$ and $d$ positive constants, \textbf{cannot} hold. This is
essentially the condition that the physical (GNS) Hamiltonian
$\tilde{H_{0}}$ is bounded from below (semibounded) (see again
\cite{Se3}), where 
$$
\tilde{H_{0}} = \sigma_{0}^{z} \sum_{k \in \mathbf{Z}^{d}} \epsilon(k) \sigma_{k}^{z}
\eqno{(40-a)}
$$
The infinite sum (40) stands for a limit in norm in the quasi-local
algebra ${\cal A}$ associated to the spin system, see  \cite{Ra},
loc. cit. It is the thermodynamic stability of the second kind (38-c)
which guarantees the existence of $\tilde{H_{0}}$, and (39) as a
consequence. In the random case (34), 
$$
\tilde{H_{0}}(J) = \sigma_{0}^{z} \sum_{k \in \mathbf{Z}^{d}} J_{0,k} \epsilon(k) \sigma_{k}^{z}
\eqno{(40-b)}
$$
under assumption (33-a): if only (33-b) holds, the representation
(40-b) is, of course, not defined. We refer to \cite{Wr5}, proposition
4.1 for the dynamical consequences of this fact: in particular, unlike
the stable case (33-a), exponential decay (39) does indeed hold for a
class of one-dimensional potentials with algebraic decay, as shown
there. 

We now consider the special nearest-neighbor interaction (35-c) -
i.e., the EA model of section 3. The following theorem may be proved
(along the same lines, but simpler, than theorem 3.2 of
\cite{Wr5}) (There are two misprints in the latter
  reference: in the r.h.s. of (40a) the number four should be inside
  the exp function, while in (40b) the first parenthesis should be
  moved to the left of the symbol Av, and the last parenthesis
  omitted):  

\textbf{Theorem 9} Let (35-c) hold, i.e., (34) describe the EA spin
glass model of section 3. Then there exists a subsequence
$\{m_{n}\}_{n \in \mathbf{Z}}$ such that, a.e. with respect to $J$, 
\begin{eqnarray*}
f(t) \equiv \lim_{m_{n} \to \infty} \rho_{\infty}^{t}(\frac{\sum_{i
    \in V_{n}}\sigma_{i}^{x}}{V_{n}})\\ 
= \delta Av(\wp_{0}(t))
\end{eqnarray*}
$$\eqno{(41)}$$
where
$$
\delta = \phi_{0}(\sigma_{0}^{x}) \ne 0
\eqno{(42)}
$$
and
$$
\wp_{0}(t) \equiv \prod_{\pm\delta_{i}}\cos(2 \beta J_{0,\pm\delta_{i}}t)
\eqno{(43)}
$$
We have the

\textbf{Corollary 9} Consider the distribution functions
$$
dP_{1}(x) = 1/2[\delta_{1}(x) + \delta_{-1}(x)] \mbox{ Bernoulli distribution } 
\eqno{(44-a)}
$$
$$
dP_{2}/dx = 1/2 \chi_{[-1,1]}(x)  \mbox{ uniform distribution } 
\eqno{(44-b)}
$$
$$
dP_{3}/dx = \frac{1}{\sqrt(\pi)} \exp(-x^{2})  \mbox{ Gaussian of unit
  variance } 
\eqno{(44-c)}
$$
The corresponding values of $f$, defined by the r.h.s. of (41), are:
$$
f_{1}(t) = (\cos(2 \beta t)^{z}
\eqno{(45-a)}
$$
$$
f_{2}(t) = (\frac{\sin (2 \beta t)}{2t})^{z}
\eqno{(45-b)}
$$
$$
f_{3}(t) = \exp(-2 z t^{2})
\eqno{(45-c)}
$$

\textbf{Remark 2}

a.) In the Bernoulli case, (45-a) implies no decay, as in the
non-random model: this is an instance of a result of Radin for general
interactions satisfying (33-a), according to which $f$ is
almost-periodic (\cite{Ra}, Lemma 2, pg.1951). 

b.) For the uniform distribution, (35-b) describes a
\textbf{non-Markovian return to equilibrium}. Clearly, in this case,
under assumption (33-a), (40b) is a physical Hamiltonian which is
semibounded for each realization of $J$, and thus exponential decay
(39) cannot hold by the aforementioned folklore theorem. 

c.) In the Gaussian case, not even thermodynamic stability of the
first kind (38-b) holds, because the Gaussian r.v. range over
$\mathbf{R}$. (40b) shows that also the physical Hamiltonian
$\tilde{H_{0}}$ is not bounded below, and thus (39) may, in principle,
hold. That this is actually the case is, of course, made clear by
(45-c). 

d.) Since the transverse magnetization in (41) is an observable (measurable quantity), self-averaging is an important requirement, as discussed before. We conjecture that all subsequential limits are equal, a.e. $J$, to the r.h.s. of (41), i.e., that the limit $n \to \infty$ in (41) exists.

\textbf{The issue of probability distributions}

As is well-known, the use of Gaussian probability distributions (p.d.)
in disordered systems is standard: it simplifies several passages
considerably, e.g., the integration by parts formula in the first
proof of the existence of the thermodynamic limit for the mean field
theory in \cite{GTo}. One might
think that the results are not expected to depend \textbf{qualitatively}
on the probability distribution: this is, however, not so. 

In \cite{SaWr}, where a mean-field Ising model in a random external
field was studied, it was proved that the existence or not of a
tricritical point in the model's phase diagram is tied to the
probability distribution: it is absent for a Gaussian p.d., and
present for a Bernoulli distribution. 

On the other hand, the results on the Anderson model on the Bethe lattice 
reviewed in 2b2, 2b3 pose absolute continuity of the probability distribution
as a major requirement. Also, as remarked by Jitomirskaya (\cite{J}, pg. 624), 
''proving localization for the Anderson model with a Bernoulli distribution
remains more than a mere technical challenge'' (this is problem 3 in Jitomirskaya
list of open problems).
 
As an attempt to explain the result in \cite{SaWr} physically, we conjectured that
it was due to the fact that a discrete distribution of probabilities
such as the Bernoulli distribution samples just a few values of the
couplings and thus introduces some short-ranged elements into the
problem: from this point of view, discrete distributions may have a
closer connection with real materials. This feature is shared by the
uniform distribution, see remark 2 b.), but even here there is a
remarkable \textbf{dynamical} difference between Bernoulli and uniform
distributions with regard to decay, exhibited by (45-a) and
(35-b). The latter yields, somewhat surprisingly, the same result in
one dimension ($z=2$) as obtained by Emch \cite{Em} with an
interaction of infinite range ($\epsilon(|n|) = 2^{-|n|-1}$) in the
nonrandom model (or, alternatively, the random model with Bernoulli
distribution and the same potential). Since the latter (algebraic slow
decay with oscillations) seemed to lead to a good qualitative
description of the nuclear spin resonance experiment \cite{LNo}, the
uniform distribution decay (45-b) might well be in qualitative
agreement with the previously proposed spin glass experiment! 

In contrast to a discrete or uniform distribution, a Gaussian
distribution samples many values of the couplings and works to
reinforce the long-range nature of the interactions. For effectively
short-range inteactions, this might not be suitable.  In any case, the
Gaussian probability distribution leads to a spectacularly fast rate
of return to equilibrium (45-c), and such sharp differences, such as
between (45-c) and (45-b), should be amenable to experiment.

\end{document}